\documentclass[twocolumn,twocolappendix]{aastex6}
\usepackage{amsmath}

\usepackage{units}

\newcommand{\me}{\ensuremath{m_{\mathrm{e}}}}
\newcommand{\Ye}{\ensuremath{Y_{\mathrm{e}}}}
\newcommand{\mprot}{\ensuremath{m_{\mathrm{p}}}}
\newcommand{\kB}{\ensuremath{k_{\mathrm{B}}}}
\newcommand{\nQ}{\ensuremath{n_{\mathrm{Q}}}}
\newcommand{\nele}{\ensuremath{n_{\mathrm{e}}}}
\newcommand{\gcc}{g\,cm^{-3}}

\newcommand{\E}[1]{\times10^{#1}}
\newcommand{\msol}{ \, M_\sun }
\newcommand{\MESA}{\texttt{MESA}}

\shorttitle{DELAYED POST-SN WINDS}
\shortauthors{SHEN \& SCHWAB}
\slugcomment{Accepted to ApJ}

\newcommand{\nuclei}[2]{\ensuremath{\mathrm{^{#1}#2}}}

\newcommand{\iron}[1][56]{\nuclei{#1}{Fe}}
\newcommand{\cobalt}[1][59]{\nuclei{#1}{Co}}
\newcommand{\nickel}[1][58]{\nuclei{#1}{Ni}}

\newcommand{\Ni}{\nickel[56]}
\newcommand{\Co}{\cobalt[56]}
\newcommand{\Fe}{\iron[56]}

\begin{document}


\title{Wait for it: Post-supernova winds driven by delayed radioactive decays}


\author{Ken J. Shen}
\affil{Department of Astronomy and Theoretical Astrophysics Center\\
 University of California, Berkeley, CA 94720, USA; kenshen@astro.berkeley.edu}

\and

\author{Josiah Schwab\altaffilmark{1}}
\affil{Department of Astronomy and Astrophysics\\ University of California, Santa Cruz, CA 95064, USA}


\altaffiltext{1}{Hubble Fellow.}

\begin{abstract}

In most astrophysical situations, the radioactive decay of \Ni\ to \Co\ occurs via electron capture with a fixed half-life of $6.1$ days.  However, this decay rate is significantly slowed when the nuclei are fully ionized because $K$-shell electrons are unavailable for capture.  In this paper, we explore the effect of these delayed decays on white dwarfs (WDs) that may survive Type Ia and Type Iax supernovae (SNe Ia and SNe Iax).  The energy released by the delayed radioactive decays of  \Ni\ and \Co\ drives a persistent wind from the surviving WD's surface that  contributes to the late-time appearance of these SNe after emission from the bulk of the SN ejecta has faded.  We use the stellar evolution code \MESA\   to calculate the hydrodynamical evolution and resulting light curves of these winds. Our post-SN Ia models conflict with late-time observations of SN 2011fe, but uncertainties in our initial conditions prevent us from ruling out the existence of surviving WD donors.  Much better agreement with observations is achieved with our post-SN Iax bound remnant models, providing evidence that these explosions are due to deflagrations in accreting WDs that fail to completely unbind the WDs.  Future radiative transfer calculations and wind models utilizing explosion simulations for more accurate initial conditions will extend our study of radioactively-powered winds from post-SN surviving WDs and enable their use as powerful discriminants among the various SN Ia and SN Iax progenitor scenarios.

\end{abstract}



\section{Introduction}
\label{sec:intro}

The radioactive decay chain of \Ni\ $\rightarrow$ \Co\ $\rightarrow$ \Fe\ contributes to the energetics of most types of supernovae and dominates the luminosity output of Type Ia, Type Iax, and Type Ib/c supernovae (SNe Ia, Iax, and Ib/c).  \Ni\ is produced in copious amounts in explosive burning that reaches nuclear statistical equilibrium because it has one of the highest binding energies per mass of any nucleus.  While \Ni\ is synthesized in extremely hot environments as a fully-ionized nucleus, the \Ni-rich material in most astrophysical situations rapidly expands and cools, and thus \Ni\ does not remain fully ionized for long.  As a result, the radioactive decay of \Ni\ occurs in the well-known way with a half-life of $6.1$ days due to $K$-shell electron captures.

However, because the decay of \Ni\ proceeds essentially 100\% via electron captures, if it remains fully ionized, its decay rate can be suppressed significantly.  In this paper, we explore the effects of this delayed radioactive decay in \Ni-rich material that may surround white dwarfs (WDs) that survive SN Ia and SN Iax explosions.  The energy released by the decay of \Ni\ through \Co\ to \Fe\ launches an optically-thick, super-Eddington wind from the WD's surface with a prolonged evolution due to the suppressed decays.  These winds will contribute significantly to the light curves and spectra of SNe Ia and SNe Iax  hundreds of days after the explosion when most of the radioactive material has disappeared from the bulk of the SN ejecta.

In \S \ref{sec:motiv}, we detail the evolutionary scenarios that result in WDs that survive SNe and derive fiducial parameters that motivate our hydrodynamical simulations.  The physics of delayed \Ni\ and \Co\ decays and our implementation of them in \MESA\ are explained in \S \ref{sec:rates}.  We describe our hydrodynamical simulations in \S \ref{sec:sim} and compare our results to late-time observations of SNe Ia and SNe Iax in \S \ref{sec:results}.  We conclude in \S \ref{sec:conc} and outline future research directions that will build on this work and help to identify the progenitors responsible for SNe Ia and SNe Iax.


\section{Motivation for Post-SN Wind Models}
\label{sec:motiv}

In this section, we describe the evolutionary scenarios in which WDs can survive SNe and capture \Ni-rich SN ejecta.  The estimates in this section motivate the initial conditions used in our fiducial models in \S \ref{sec:results}.


\subsection{Type Ia Supernovae and Surviving WD Donors}
\label{sec:Iamotiv}

The identity of the progenitors of Type Ia supernovae (SNe Ia) remains an unsolved mystery (\citealt{maoz14a} and references therein).  One of the most promising scenarios at present invokes helium-rich accretion from a helium WD or low-mass C/O WD.  If mass transfer is dynamically stable, the resulting convective burning episodes may eventually yield a helium shell detonation \citep{bild07,sb09b} and a subsequent core explosion in a variant of the double detonation scenario \citep{woos80a,fhr07,shen14a}.  For systems that merge unstably, which may in fact constitute most double WD binaries \citep{shen15a}, dynamical instabilities caused by the interaction of the direct impact accretion stream with freshly accreted material may trigger a helium detonation during the merger \citep{guil10,dan12,rask12,pakm13a}, also yielding  a double detonation SN Ia.

In both of these cases, the WD donor may survive the explosion of the accretor.  In the dynamically stable mass transfer case, the donor is only mildly tidally distorted as it overflows its Roche lobe and will remain self-bound following the SN Ia.  For dynamically unstable double WD mergers, the ease with which helium polluted with CNO nuclei can be detonated \citep{shen14b} allows for the possibility that a helium detonation is triggered soon after unstable mass transfer begins in helium-rich double WD binaries.  If this is the case, the primary WD may explode and produce a SN Ia before the donor is tidally disrupted.\footnote{\cite{papi15a} find that helium WD companions can be detonated and destroyed by the impacting SN Ia ejecta.  However, this is only the case when the companion WD is placed artificially closer than its separation at Roche lobe overflow.  The companion WD survives if at the appropriate distance from the center of the explosion.}

The companion WD may survive the SN Ia explosion, but it will not emerge unchanged.  The focus of this work is the \Ni\ that is synthesized by the primary WD's explosion at velocities low enough such that the \Ni\ is gravitationally bound to and captured by the WD companion.  We  estimate the bound \Ni\ mass using \cite{dwar98a}'s approximation to the homologously expanding SN Ia ejecta density profile,
\begin{equation}
	\rho(v,t) = \frac{M_{\rm ej} }{ 8 \pi  v_{\rm ej}^3 t^3} \exp \left( -v/v_{\rm ej} \right) ,
\end{equation}
where $M_{\rm ej}$ is the ejecta mass, $t$ is the time from explosion, $E_{\rm kin}$ is the ejecta's total kinetic energy, $v_{\rm ej}^2 = E_{\rm kin} / 6 M_{\rm ej}$, and $v$ is the magnitude of the velocity in the SN's center of mass frame.  Ignoring the work that is done on the ejecta by the shocked material interacting with the companion, the mass bound to the companion is
\begin{equation}
	M_{\rm bound} =  \int \rho(\vec{v},t) dV
\end{equation}
such that $ (\vec{v}-\vec{v}_2)^2 < 2 G M_2 / d$, where $v_2= \sqrt{G (M_1+M_2)/a}$ is the magnitude of the companion's velocity in the SN ejecta's rest frame, $a$ is the binary separation at the time of explosion, $d$ is the distance from the material to the companion's surface,  and $M_2$ is the companion's mass.  Using the explosion values from \cite{sim10} for a  $M_{\rm ej} = 0.97 \msol$ pure detonation explosion with $1.04 \times 10^{51}$ erg of kinetic energy, a mass-radius relation for $10^7$~K WDs from \MESA\ \citep{Paxton11, Paxton13, Paxton15}, and \cite{pacz71}'s fit to the Roche radius, and evaluating at $t= \unit[1]{s}$, we find bound \Ni\ masses of $0.006$, $0.03$, and $0.08 \msol$ for $0.3$, $0.6$, and $0.9 \msol$ WD companions, respectively.

We note that these are upper limits to the actual bound masses, because we have neglected the work that is done on the ejecta by previously shocked material surrounding the companion.  Furthermore, the surface structure of the surviving companion WD will be affected by the passage of the shock and by the Roche lobe overflow that takes place prior to the SN Ia explosion, particularly for merger-induced detonations.   These effects will  influence the amount of \Ni\ that is captured by the surviving WD donor, as well as contributing to the light curve when the thermalized shock energy is radiated outwards. We parametrize these uncertainties in \S \ref{sec:results} by varying the bound envelope masses, but we emphasize that future work with more realistic initial conditions is necessary for more accurate results.


\subsection{Type Iax Supernovae and Surviving WD Accretors}
\label{sec:Iaxmotiv}

Type Iax supernovae (SNe Iax) are a recently designated class of peculiar SNe  with characteristics akin to the prototype SNe 2002cx \citep{li03} and 2005hk \citep{phil07}.  While some debate remains \citep{vale09,mori10a}, the current observational evidence suggests that they are thermonuclear explosions of WDs with non-degenerate donor stars (\citealt{fole13a} and references therein).  In one case, SN 2012Z, the progenitor system was detected and is consistent with a helium-burning star donor \citep{mccu14a}.

The prevailing SN Iax progenitor scenario (e.g., \citealt{fole13a}) is very similar to the formerly standard SN Ia progenitor scenario (e.g., \citealt{nty84}): a WD accretes matter from a non-degenerate donor, grows towards the Chandrasekhar mass, and ignites nuclear burning in its core due to the increasing density.  The energy release from carbon-burning leads to a $\sim \unit[1000]{yr}$ convective simmering phase \citep{wwk04,pc08} that ends when turbulent fluctuations give rise to hotspots that heat faster than convective eddies can dissipate their excess entropy, resulting in the birth of a deflagration.   The important difference between a SN Ia and a SN Iax outcome is whether or not this deflagration transitions to a detonation.  Without this transition, the buoyant deflagration plume rises and breaks out of the star, burning the WD incompletely and not releasing enough energy to unbind the entire WD.  The ejecta mass and the ejected \Ni\ mass are thus smaller than for a regular SN Ia, leading to fainter and more rapidly-evolving light curves.

A very important consequence of this model is the surviving WD remnant and the deflagration ashes that remain bound to this remnant.  Simulations show that much of the deflagration ash is ejected \citep{jord12a,krom13a,fink14a,long14a}.  However, a significant fraction of the WD remains unburned and gravitationally bound following the ejection of the burned material, and this surviving WD has a large enough potential well to retain $0.03-0.2 \, M_\odot$ of burned material as a gravitationally bound envelope.  Radioactive decays in this burned material will drive a wind that neatly explains the persistent photosphere, low velocities, and lack of nebular emission that are some of the hallmarks of SNe Iax \citep{jha06a,fole13a}.  As in the case of the surviving SN Ia WD donors discussed in \S \ref{sec:Iamotiv}, the dense, hot \Ni-rich matter at the surface of the bound SN Iax WD accretor remnant will be fully ionized, leading to delayed radioactive decays and a prolonged WD wind.

We use the results from \cite{fink14a} to motivate our SN Iax WD wind models.  They performed a large suite of simulations with a varying number of deflagration ignition kernels.  Their best fits to the prototype SN 2005hk were for models with $5-20$ ignition kernels, resulting in unburned masses of $0.91-0.49 \msol$ and bound, burned masses of $0.12-0.06 \msol$ containing \Ni\ masses of $0.02-0.004 \msol$.  These numbers are in rough agreement with the other simulations in the literature \citep{jord12a,krom13a,long14a}.   We bracket the uncertainties by considering two models with core $+$ envelope masses of $0.9+0.1$ and $0.5+0.05 \msol$ and \Ni-mass fractions within the envelopes of $0.2$ and $0.1$, respectively.  The non-\Ni\ material in the envelope is assumed to be \Fe, because it is radioactively inert.

One complication is the possibility that, following the ejection of material, the bound WD remnant continues to convectively burn carbon in its core, which would eventually lift the degeneracy of the core and significantly impact the subsequent evolution.  However, the ejection of a large amount of material may yield enough adiabatic expansion to quench further convective core burning.  We test this by constructing a near-Chandrasekhar mass $1.35 \msol$ C/O WD in \MESA\ and heating the center until carbon-burning causes convection.  We find that a localized burning region begins to run away in the center when the convective region's entropy is $ \unit[8\E{7}]{erg\,  g^{-1}\, K^{-1}}$.  We then construct less massive C/O stars with the same entropy, approximating the SN Iax WD remnants after the ejection of the burned material, and follow their evolution to ascertain whether or not they continue to burn convectively.  We find that even for a remnant mass of $1.2 \msol$, which is larger than we consider in this paper, the removal of the inner $0.15 \msol$ from the simmering near-Chandrasekhar mass WD decreases the gravitational potential sufficiently that the remnant's center cools without reigniting convective carbon-burning.  We are thus justified in assuming that the bound SN Iax remnants we consider are simply degenerate WD cores with hot \Ni-rich envelopes.


\section{Decay Rates}
\label{sec:rates}

We now turn to an analysis of fully-ionized \Ni\ and \Co\ decays and
our implementation of the resulting decay rates in \MESA.  Electron
capture is the primary decay channel for both \Ni\ and \Co\ under
laboratory conditions \citep[e.g.,][]{Nadyozhin94}.  For \Ni\ and \Co\
with bound electrons, the electron density at the nucleus is dominated
by the $K$-shell electrons.  It is the capture of these electrons that
sets their familiar $\unit[6.1]{day}$ (\Ni) and $\unit[77]{day}$ (\Co)
half-lives.  Decay of fully-ionized \Ni\ or \Co\ in an environment
effectively free of electrons can still proceed by positron emission,
but this channel is $\gtrsim 10^6$ (\Ni) or $\approx 5$ (\Co) times
slower than electron capture \citep{Sur90, daCruz92}.  Such an extreme
limit is not relevant for dense stellar interiors, where \Ni\ and \Co\
nuclei can still capture an electron from the continuum.  However, for
densities $\lesssim \unit[10^5]{\gcc}$, the continuum electron density
is lower than the effective density of the $K$-shell electrons.
Therefore, the decays of fully-ionized atoms are less rapid than those
with bound electrons.  The delayed decays of \Ni\ and \Co\ have
been previously discussed in the context of cosmic rays
\citep{LundFisker99} and high-velocity ejecta from gamma-ray bursts
\citep{McLaughlin02}, and the analogous delay for $^{44}$Ti decays in
young SN remnants was studied by \cite{moch99a}.

The hot \Ni\ retained by post-SN surviving WDs is at
thermodynamic conditions where such considerations are
important.  The existing treatment of these rates in \MESA\
is correct in both the hot, high density limit (where atoms are
assumed to be fully ionized and the tabulated rates of
\citealt{Langanke00} are used) and the cool, low density limit (where
the atoms are assumed to have at least their inner electrons and the
``lab'' rates compiled in \citealt{Nadyozhin94} are used).  However,
much of the material of interest in the surface layers of the
surviving WDs is in the regime between these two limits.

\begin{figure}
  \centering
  \includegraphics[width=\columnwidth]{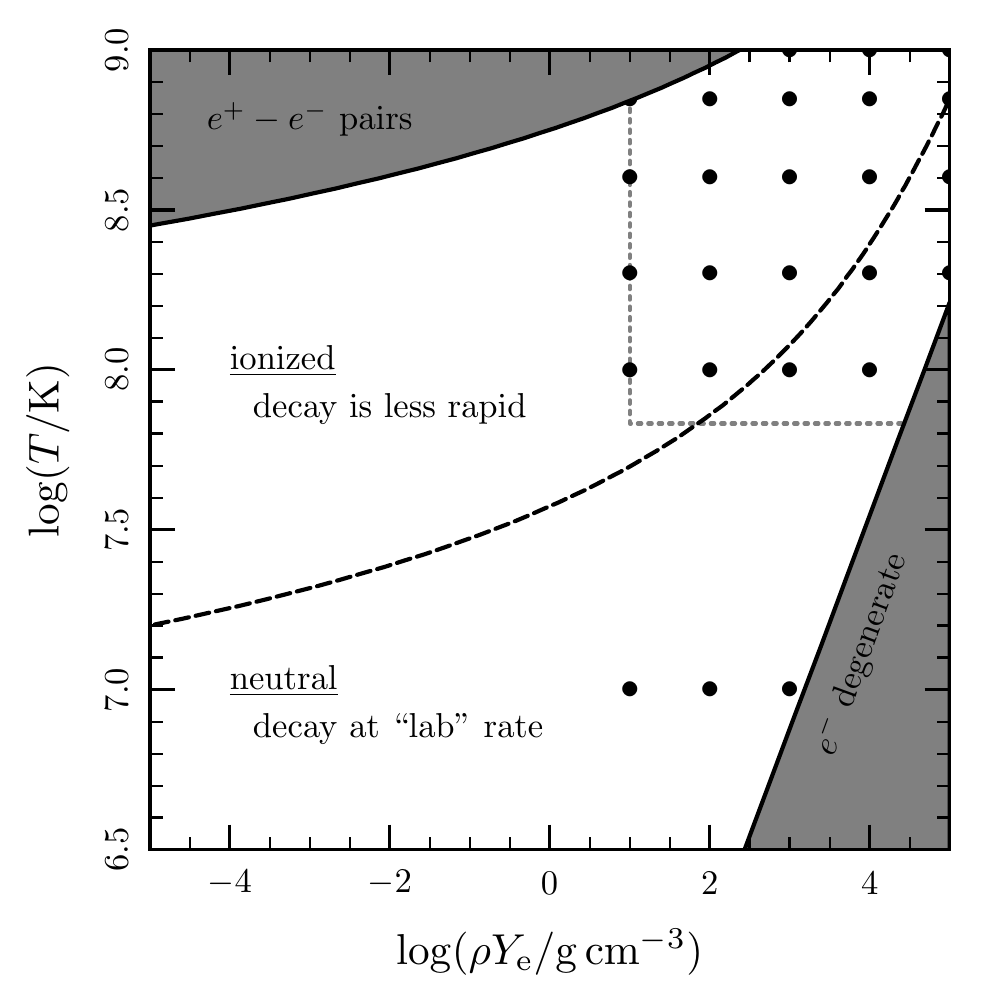}
  \caption{Regimes in density-temperature space for  \Ni\ and \Co\
    decays.  The shaded region in the upper left indicates where the
    electron density becomes dominated by electrons from pair
    production; the shaded region in the lower right shows where the
    electrons become degenerate.  The black circles mark the points in
    the rate tabulations of \citet{Langanke00} that are used in \MESA.
    The dotted lines delineate the region of the table that \MESA\
    uses by default for reactions on isotopes with $Z \ge 26$; this
    restriction represents a rough approximation of where these
    elements are fully ionized.  The dashed line shows the location of
    half-ionization for the $K$-shell electrons
    (eq.~\ref{eq:half-ionized}).  Above this line, where the \Ni\ or
    \Co\ is ionized, the average electron density at the nucleus will
    be lower and hence the decay rate can be significantly slower than
    the ``lab'' rate.}
  \label{fig:decay-regimes}
\end{figure}

Figure~\ref{fig:decay-regimes} summarizes the regimes in
density-temperature space.  The dashed line shows an approximate boundary
for the half-ionization state of the $K$-shell electrons.  Below this line,
the decay rate will be roughly the lab rate.  Above this line, the
rate will decrease as the fraction of fully-ionized atoms increases.


\subsection{Ionization State}

In order to determine the ionization state of the material, we solve
the two Saha equations for the occupation fractions of the
fully-ionized $(f_0)$, hydrogen-like $(f_1)$, and helium-like states
$(f_2)$.\footnote{Note that we label our
  occupation fractions by the number of bound electrons, as opposed to
  the ion charge (e.g., 0 represents fully-ionized and not neutral).}
We do not need to track lower-ionization states since the electron
density at the nucleus is primarily determined by the $K$-shell.
Therefore we have
\begin{align}
  \frac{f_0}{f_1} &= \frac{g_{\mathrm{e}} g_0}{g_1} \left(\frac{\nQ}{\nele}\right) \exp\left(-\frac{\chi_1}{\kB T}\right)\\
  \frac{f_1}{f_2} &= \frac{g_{\mathrm{e}} g_1}{g_2} \left(\frac{\nQ}{\nele}\right) \exp\left(-\frac{\chi_2}{\kB T}\right)
\end{align}
along with the constraint $f_0 + f_1 + f_2 = 1$.  The degeneracies are
$g_{\mathrm{e}} = g_1 = 2$, $g_0 = g_2 = 1$, and the quantum concentration is
 $\nQ = \left(  \me \kB T / 2\pi\hbar^2 \right)^{3/2}$.  The values
$\chi_1$ and $\chi_2$ are the ionization energies of these states,
which are drawn from \citet{NIST_ASD} and are listed in
Table~\ref{tbl:nico-info}.

\begin{deluxetable}{ccccc}
\tablenum{1}
\tablecaption{Information about \nickel[56] and \cobalt[56]\label{tab:rates}}
\tablewidth{0pt}
\tablehead{
\colhead{Reaction} & \colhead{``Lab'' lifetime} & \colhead{$Q_\nu$} & \colhead{$\chi_1$} & \colhead{$\chi_2$} \\
\colhead{} & \colhead{(s)} & \colhead{(MeV)} & \colhead{(keV)} & \colhead{(keV)}
}
\startdata
$\Ni\ \to \Co$ & $1.32 \times 10^{-6}$ & 0.41 & 10.775 & 10.289 \\
$\Co\ \to \Fe$ & $1.04 \times 10^{-7}$ & 0.84 & 10.012 & 9.544 \\
\enddata

\begin{minipage}{\columnwidth}

  \tablecomments{The values of the ``lab'' rate and $Q_\nu$ (the
    average energy of the emitted neutrino) are from
    \citet{Nadyozhin94}. The values $\chi_1$ and $\chi_2$ are the
    ionization energies of the hydrogen-like and helium-like states,
    respectively, and are drawn from \citet{NIST_ASD}.}

\end{minipage}
\label{tbl:nico-info}
\end{deluxetable}

Furthermore, we make the assumption that the electron density is
independent of the $f$ values.  If the \Ni\ is a trace isotope, this
is essentially exact.  In the opposite limit of pure \Ni, this is
justified since we are considering only the highest ionization states:
if all \Ni\ goes from fully ionized to having two bound electrons, the
electron density changes by a fraction $\approx 2/Z$ which is only
$\approx 7\%$.  We use the electron density reported by \MESA,
which is for a fully-ionized composition, i.e.,
$\nele = \rho \Ye/\mprot$.  With these occupation fractions, the average
number of bound electrons is equal to 1 when
\begin{equation}
  \label{eq:half-ionized}
  \nele = 2 \nQ \exp\left(-\frac{\chi_1 + \chi_2}{2 \kB T}\right).
\end{equation}
This is the dashed line shown in Figure~\ref{fig:decay-regimes}, where
the $\chi$ values used are those of \Ni.

\begin{figure}
  \centering
  \includegraphics[width=\columnwidth]{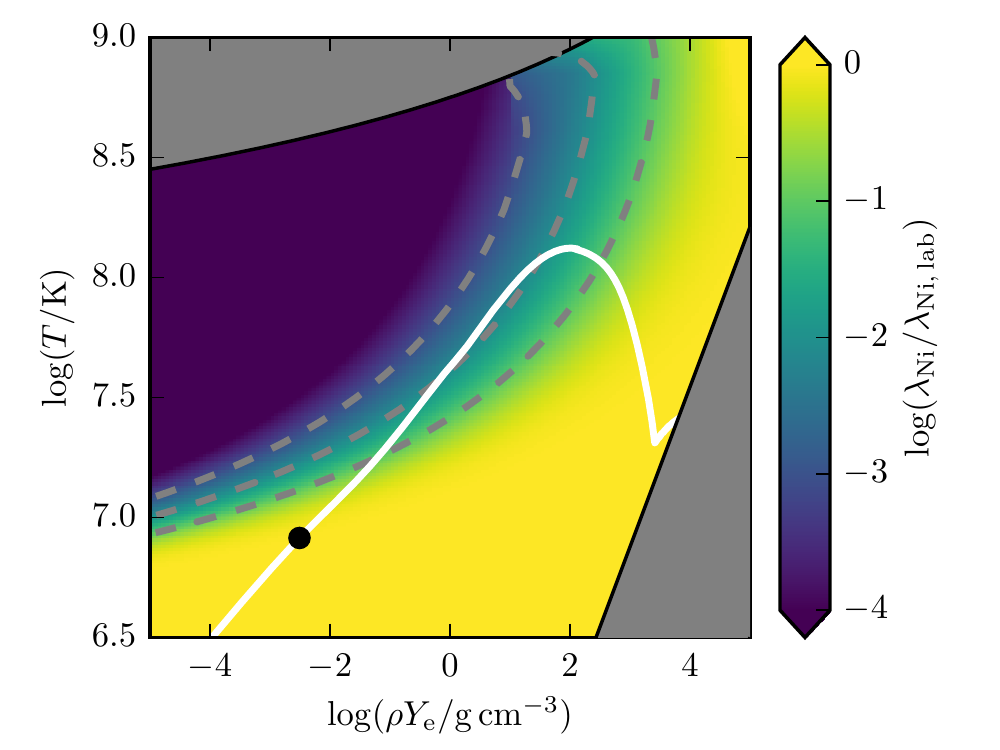}
  \includegraphics[width=\columnwidth]{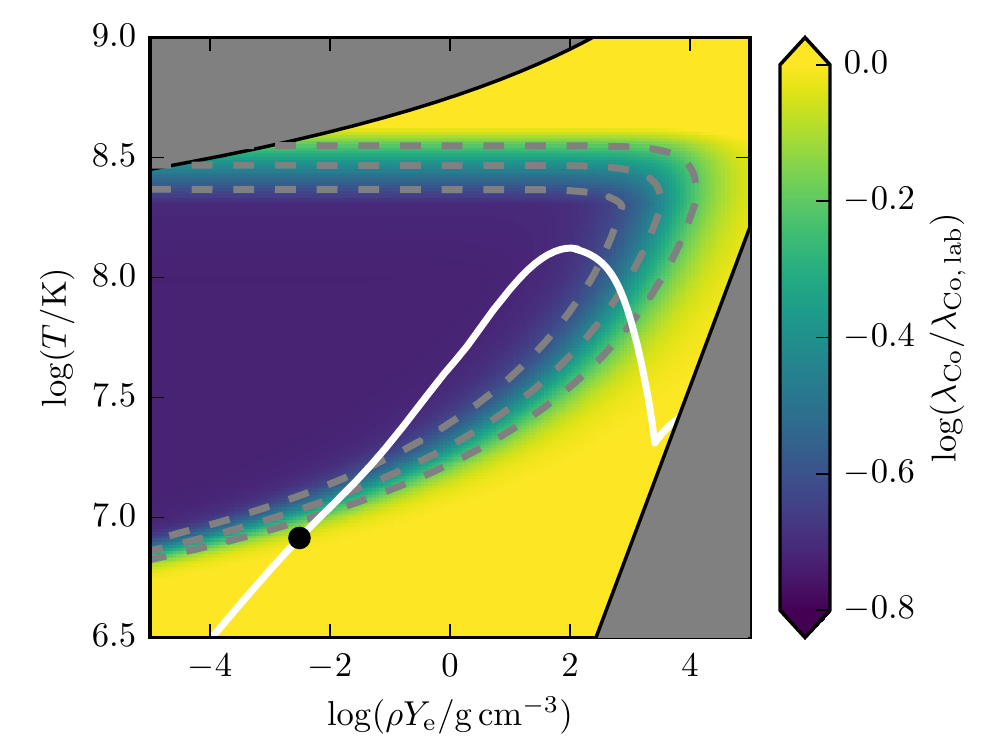}
  \caption{Rate of \Ni\ decay (top plot) and \Co\ decay (bottom plot)
    used in this work.  The rate is shown relative to the ``lab''
    rate.  The gray shaded regions are the same as in
    Figure~\ref{fig:decay-regimes}. Dashed contours are shown at the
    values of the tick marks in the colorbars.  Note the much smaller
    dynamic range in the case of \Co\ compared to \Ni. The solid
    white line shows the density-temperature profile of the main fiducial post-SN Ia
    model $ \unit[500]{d}$ after the explosion, as discussed in \S~\ref{sec:results}; the black dot on
    this profile indicates the conditions at the location of the
    maximum nuclear energy generation rate.}
  \label{fig:mesa-rates}
\end{figure}

\subsection{Extension of \texttt{weaklib} Tables}

As shown in Figure~\ref{fig:decay-regimes}, the tabulated
\texttt{weaklib} rates do not extend to densities lower than $\rho Y_e = \unit[10]{g\, cm^3}$.  Therefore, in order to construct smooth decay rates
for the fully-ionized atoms, we make the following simple extensions
of these tables.

In the case of \Ni, since the positron decay rate is so slow, we only need
 to consider electron capture from the continuum.  The
electron-capture rate corresponding to a single transition can be
written as
\begin{equation}
  \label{eq:lambda-ec}
  \lambda_{\mathrm{ec}} \propto \int_{1}^{\infty} \frac{\epsilon^2(\epsilon+q)^2}{1 + z^{-1} \exp\left(\theta \epsilon\right)} G(Z,\epsilon) d\epsilon
\end{equation}
with the usual thermodynamic definitions $\beta = (k_{\mathrm{B}} T)^{-1}$,
$\theta = \beta \me c^2$, and $z = \exp(\beta \mu_{\mathrm{e}})$, where  $T$ is
the temperature and $\mu_{\mathrm{e}}$ is the electron chemical potential \citep{bahc64}.  We have
also defined a dimensionless energy difference $q = Q / (\me c^2)$ and
a dimensionless energy for the electron,
$\epsilon = E / (\me c^2)$.  In the non-relativistic limit,
$G(Z,\epsilon) \approx 2 \pi \alpha Z$ (equation 5c in
\citealt{Fuller80}) where $\alpha$ is the fine structure constant.  In
the non-degenerate limit, we can replace the Fermi-Dirac distribution
with a Boltzmann distribution where $z = (\nele/\nQ) \exp(\theta)$.
Thus, to leading order in $\theta$,
\begin{equation}
  \lambda_{\mathrm{ec}} \propto \frac{1}{\theta}\frac{\nele}{\nQ} \propto \rho T^{-1/2}.
\end{equation}   
Therefore, when we are off the table, we use the rate
\begin{equation}
  \lambda_{\mathrm{off,Ni}}(\rho\Ye,T) =
  \left(\frac{\rho\Ye}{\unit[10]{\gcc}}\right)
  \left(\frac{T}{\unit[10^7]{K}}\right)^{-1/2}
  \lambda_{\mathrm{ref,Ni}}
\end{equation}
where $\lambda_{\mathrm{ref,Ni}}$ is the \texttt{weaklib} rate at
$\rho Y_e = \unit[10]{g\, cm^{-3}}$ and $T = \unit[10^7]{K}$.

In the case of \Co, at densities below those included in
\texttt{weaklib} tables, the dominant decay for fully-ionized material
is positron emission. The
positron-emission rate corresponding to a single transition can be
written as
\begin{equation}
  \label{eq:lambda-pe}
  \lambda_{\mathrm{pe}} \propto \int_{1}^{q} \epsilon^2(\epsilon-q)^2 G(-Z,\epsilon) d\epsilon
\end{equation}
with the symbols having the same definitions as in
equation~\eqref{eq:lambda-ec}.  We have assumed that positron phase
space is empty. In the non-relativistic limit,
$G(-Z,\epsilon) \approx 2 \pi \alpha Z \exp(-2\pi\alpha Z \epsilon /
\sqrt{\epsilon^2-1})$ (equation 5d in \citealt{Fuller80}).  Therefore
equation~\eqref{eq:lambda-pe} is simply a constant, and thus the rate
has no density or temperature dependence. This is only true for a
single transition.  A significant temperature dependence can still
exist, and does in this case, due to the thermal occupation of the first
nuclear excited state at $E = \unit[0.158]{MeV}$, allowing additional
positron-decay transitions to become important.  However, this does
not introduce a density dependence.  Therefore, when we are off the
table, we use the rate
\begin{equation}
  \lambda_{\mathrm{off,Co}}(\rho\Ye,T) = \lambda_{\mathrm{ref,Co}}(T)
\end{equation}
where $\lambda_{\mathrm{ref,Co}}(T)$ is the \texttt{weaklib} rate at
$\rho Y_e= \unit[10]{g\, cm^{-3}}$.

\subsection{Rates Used in \MESA}

We use the ionization fractions (\S 3.1), ``lab'' rates (Table 1), and
the extended \texttt{weaklib} rates tables (\S 3.2) to construct rates
that are applicable over the regime of interest.  We assume that atoms
with two (or more) bound electrons decay at the ``lab'' rate and that
atoms with exactly one bound electron decay at one-half of the lab
rate. Atoms that are fully ionized decay at the rate given by the extended
\texttt{weaklib} tables.  Therefore, the total decay rate is
\begin{equation}
  \label{eq:total-rate}
  \lambda_{\mathrm{total}} = (f_2 + 0.5 f_1) \lambda_{\mathrm{lab}} + f_0 \lambda_{\mathrm{weaklib}}~.
\end{equation}
The rates that we use in our \MESA\ calculations are shown in
Figure~\ref{fig:mesa-rates}.


\section{Simulation Setup}
\label{sec:sim}

With the initial conditions from \S \ref{sec:motiv} and the modifications to the radioactive decays from \S \ref{sec:rates} in hand, we are now ready to describe our post-SN WD wind calculations.  These hydrodynamic calculations are performed with the stellar evolution code \MESA\footnote{http://mesa.sourceforge.net, version 8118; default options used unless otherwise noted.} \citep{Paxton11, Paxton13, Paxton15}.  We describe some important modifications to the default options here; an example inlist with the complete set of changed parameters can be found in the Appendix.

Our initial models consist of degenerate cores  with equal mass fractions of $^{12}$C and $^{16}$O with hot, constant-entropy \Ni-rich layers on top.  Our fiducial models have envelope entropies twice that for which the envelope's total internal energy is equal to half the envelope's gravitational binding energy.  This very roughly approximates the configuration of  a loosely-bound \Ni-rich envelope that would be captured by a surviving WD donor from a SN Ia or would fall back onto the WD accretor after a SN Iax.  We compare the effects of varying the entropy in \S \ref{sec:results}, but we acknowledge that utilizing realistic entropy profiles is an important next step for future work.

The degenerate cores are cooled to a constant temperature of $ \unit[3 \times 10^7]{K}$.  For models with very low-mass envelopes or cores, the core's outer layers will not be degenerate at this temperature.  In these cases, the outer edge of the core is set to a constant entropy of $\unit[2 \times 10^8]{erg\, g^{-1}\,K^{-1}}$.

Once our initial models are constructed, we turn on \MESA's hydrodynamic capabilities.  We also make use of recently implemented outer boundary conditions that were developed for modeling super-Eddington stellar winds \citep{quat16a}.  These allow us to set the optical depth at the outer boundary to be $\tau=10^{-4} \times 2/3$, so that we can follow the wind out to arbitrarily large distances.

We use a simple nuclear network that adds $^{28}$Si, \Fe, \Co, and \Ni\ and their interlinking reactions to the default \texttt{basic.net}.  Our modifications to the decay rates of \Co\ and \Ni\ are described in \S \ref{sec:rates}.  The energy released by radioactive decays above the $\tau=2/3$ photosphere should not contribute to the luminosity structure of this optically-thin region.  We thus neglect the nuclear energy released in optically-thin regions during the simulations, although we do add the power from the thermalization of positron kinetic energy to the total luminosity in the figures in \S \ref{sec:results}.

We treat the opacity in a simplified manner by setting it to a constant value of $\kappa = \unit[0.2]{cm^2\,g^{-1}}$ in material with an iron group mass fraction $X_{56} >0.1$ and a density $\rho <\unit[1]{\gcc}$, because \MESA\ opacity tables do not extend to our iron-group-dominated compositions of interest.  More practically, this avoids issues with very sharp opacity gradients in the \MESA-provided tables in some regions of very low density and temperature.  We perform comparison simulations with $\kappa = \unit[2.0]{cm^2\,g^{-1}}$, which we show in \S \ref{sec:results}.  At the end of some simulations, when the WD evolves to the cooling track after several years or decades, our simplified treatment causes an abrupt change in opacity at $\rho=\unit[1]{\gcc}$ that also leads to timestep problems.  In these situations, we remove the constant opacity constraint and use the \MESA\ tables.  We leave a proper treatment of iron-group-dominated opacity, including line-broadening effects in the accelerated wind, to future work.

We constrain convective velocities to be less than the local sound speed and limit the acceleration of these velocities to the local gravitational acceleration.  Furthermore, we fully switch off convection in the first part of the simulation until a wind has become established or until three days have elapsed if a wind does not occur before then.  This avoids the excessively small timesteps that may occur due to convection-induced shocks during the establishment of the wind.

Once the luminosity generated within the envelope drops below the Eddington limit, the wind ceases, and material near the acceleration region begins to fall back onto the WD.  The spatial resolution decreases significantly in this region, causing the timestep to become very short.  Since the wind-launching phase of the evolution is completed by this point, we remove the unbound mass above the photosphere rapidly, within several hundred timesteps.  The outer $\sim 10^{-7} \msol$ of the remaining star is perturbed by this procedure, but the thermal time through this layer is $\sim 2 $ d, much shorter than the timescales of interest, so we are justified in performing this manipulation of the stellar model.


\section{Results and Comparisons to Observations}
\label{sec:results}

In this section, we show the results of a suite of WD wind simulations, encompassing both surviving SN Ia WD donors and SN Iax bound WD accretor remnants.


\subsection{Post-SN Ia Winds}

\begin{figure}
  \centering
  \includegraphics[width=\columnwidth]{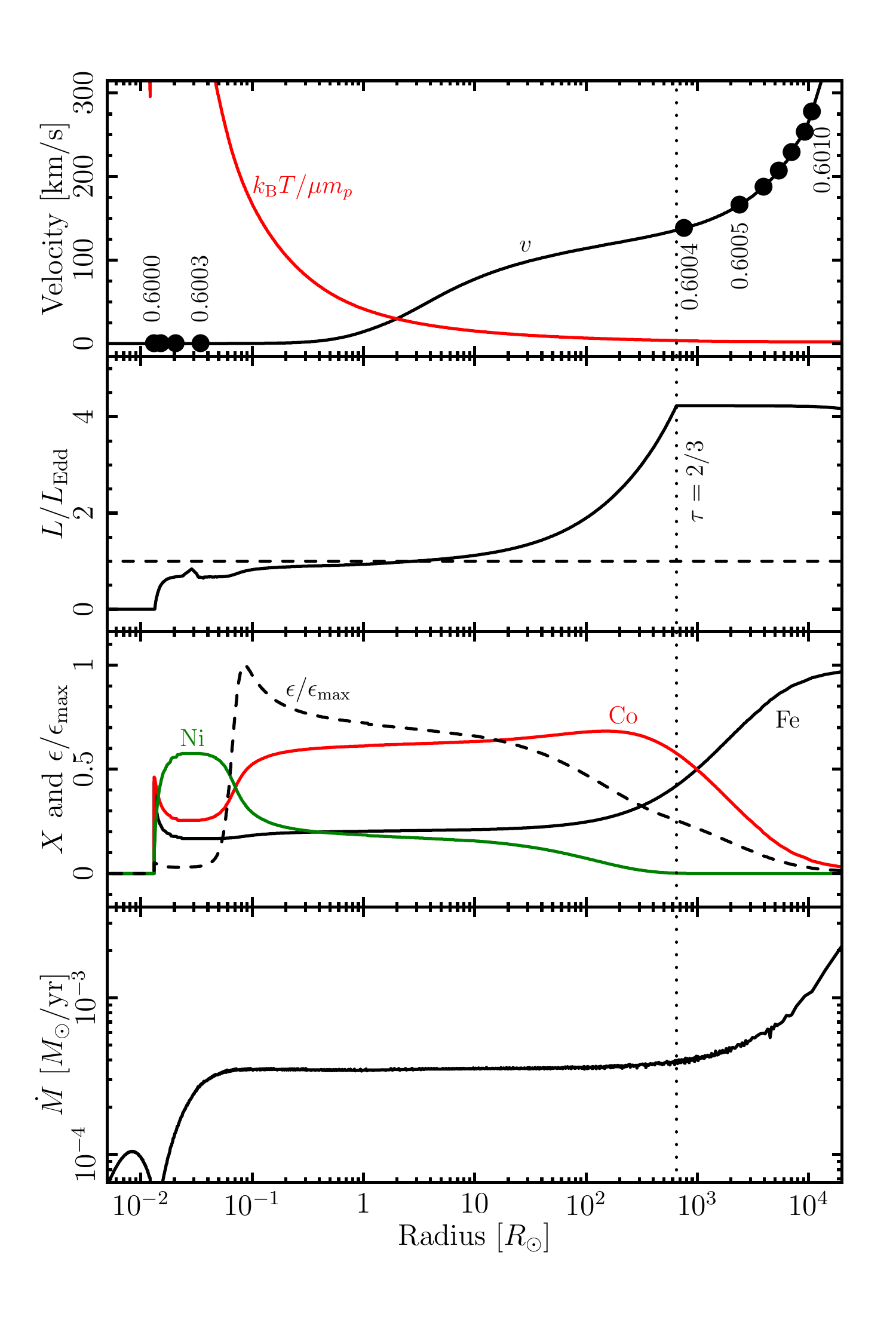}
  \caption{Profiles as a function of radius at $ \unit[500]{d}$ for our main fiducial post-SN Ia model of a $0.6 \msol$ core and a $0.03 \msol$ \Ni\ envelope.  The top panel shows the velocity (black) and isothermal sound speed (red), with bullets demarcating enclosed masses separated by $10^{-4} \msol$.  The second panel shows the ratio of the luminosity to the Eddington luminosity, $L_{\rm Edd}=4 \pi G m c / \kappa$ (solid); the dashed line shows a ratio of one. The third panel shows the mass fractions of \Ni\ (green), \Co\ (red), and \Fe\ (black), and the nuclear energy generation rate, $\epsilon$, normalized to the maximum value $\epsilon_{\rm max}$. The mass outflow rate, $\dot{M}$, is shown in the fourth panel.  The dotted line marks the location of the $\tau=2/3$ photosphere in all the panels.}
  \label{fig:profiles}
\end{figure}

Figure \ref{fig:profiles} shows profiles at $500$ d of our main fiducial SN Ia surviving companion model, a $0.6 \msol$ WD that has captured a $0.03 \msol$ envelope that is initially pure \Ni.  The radii of various enclosed masses, separated by $10^{-4} \msol$, are shown as bullets, some of which are labeled with vertical numbers.  The third panel shows the mass fractions of \Ni\ (green), \Co\ (red), and \Fe\ (black); a significant amount of undecayed \Ni\ remains even after 82 ``lab'' half-lives have elapsed.  Due to the steady wind (as demonstrated in the bottom panel by the constant value of $\dot{M}=3.5 \times 10^{-4} \, M_\odot \,$yr$^{-1}$ within the acceleration region), the part of the envelope near $0.1 \, R_\odot$ has expanded to a density and temperature where \Ni\ is no longer fully ionized and electron captures can occur.  At this location, the energy generation rate reaches a maximum (third panel, dashed line), which is also seen in the density-temperature profiles in Figure \ref{fig:mesa-rates}. As expected from steady-state super-Eddington wind solutions \citep{nugi02a,ro16a}, the location where the luminosity surpasses the local Eddington rate (second panel) is nearly coincident with the sonic point with respect to the isothermal gas sound speed, $k_BT/\mu m_p$, where $\mu$ is the mean molecular weight (top panel).

Due to the continued energy release from \Ni\ and \Co\ decays, the luminosity and velocity increase above the wind-launching region.  The luminosity increase ends at the $\tau=2/3$ photosphere (dotted line in all panels), because the energy from radioactive decays is assumed to escape this optically-thin region.  To be more precise, the kinetic energy from the positrons is likely thermalized even in this optically-thin region.  We account for this small effect in the light curves shown later in this section, but we neglect this energy when evolving the models with \MESA.  The velocity and mass loss rate, however, continue to increase above the photosphere.  This is not due to continual acceleration, but instead to previously higher velocities and higher mass loss rates earlier in the model's evolution.

\begin{figure*}
  \centering
  \includegraphics[width=0.45\textwidth]{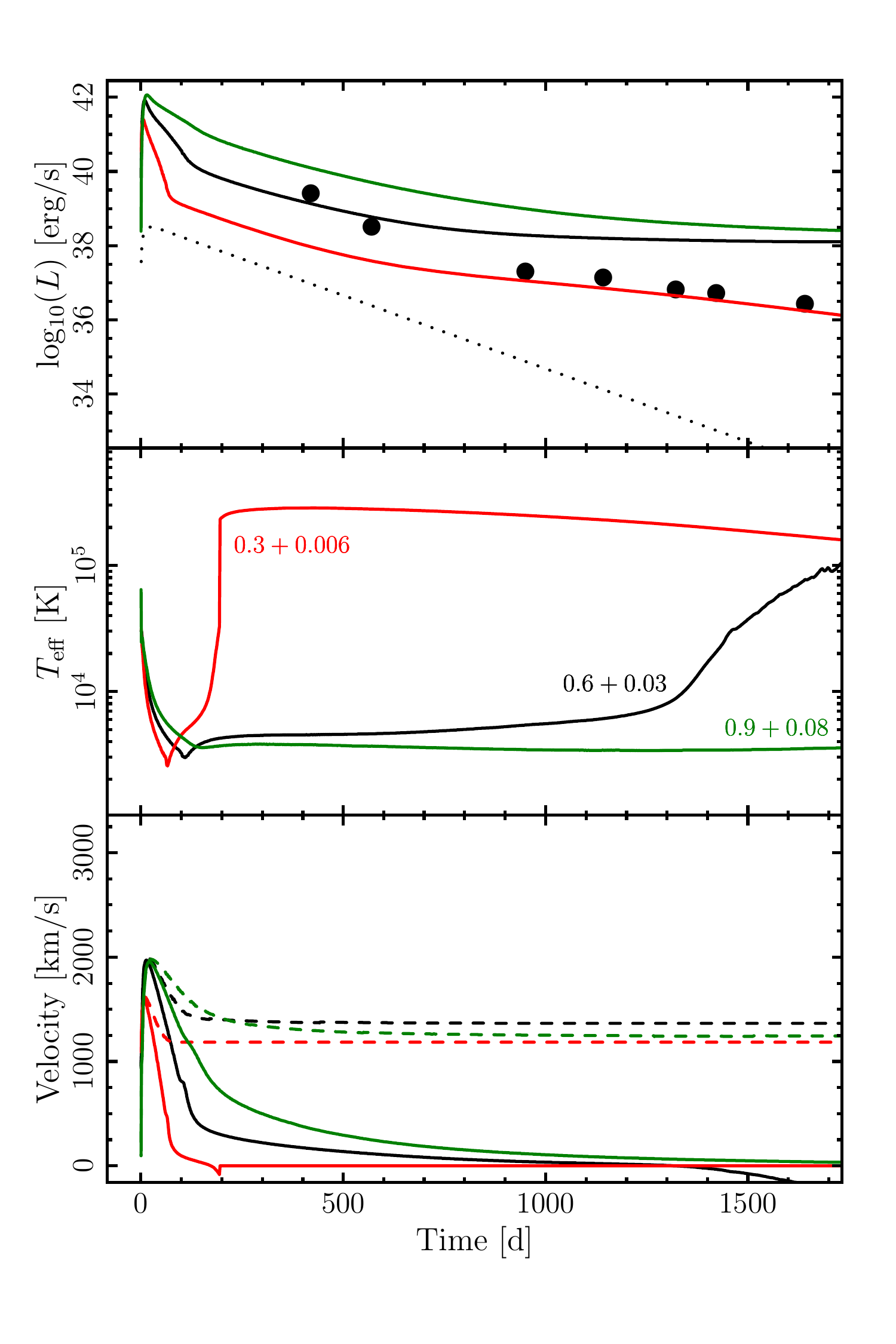}
  \includegraphics[width=0.45\textwidth]{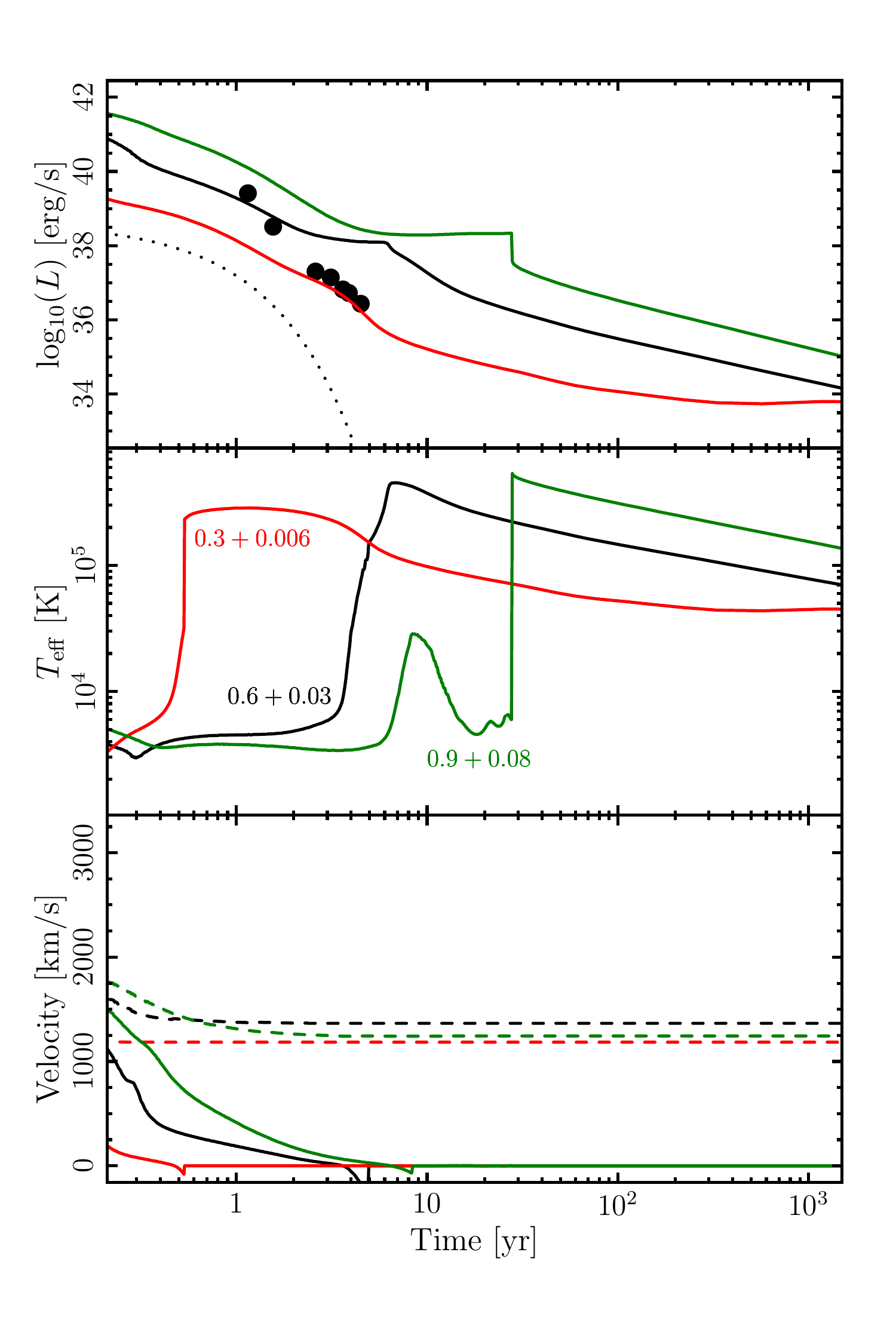}
  \caption{Evolution of winds launched from post-SN Ia surviving WD donors with envelope masses determined by the maximum bound ejecta mass.  Three different models are shown, with core + envelope masses of $0.6+0.03$ (black), $0.3+0.006$ (red), and  $0.9+0.08 \msol$ (green).  The left-hand and right-hand plots use linear and logarithmic time axes, respectively.  The top panel in each plot shows the sum of the luminosity at the $\tau=2/3$ photosphere and the luminosity from thermalization of positron kinetic energy from \Co\ decays above the photosphere.  The dotted line represents the total power from radioactive decays of $0.001 \msol$ of initially pure \Ni\ under ``lab'' conditions.  Note that the luminosity from the SN Ia ejecta is not included.  The middle panels track the effective temperature at the $\tau=2/3$ location.  The bottom panels show the velocity at the $\tau=2/3$ photosphere (solid) and the mass-averaged velocity of material above the photosphere (dashed).}
  \label{fig:wdtest}
\end{figure*}

In Figure \ref{fig:wdtest}, we show the linear (left-hand plot) and logarithmic (right-hand plot) temporal evolution of our main fiducial model as well as fiducial models with different WD and envelope masses.  Black, red, and green lines represent core + envelope masses of $0.6+0.03$, $0.3+0.006$, and $0.9+0.08 \msol$, respectively.  In each plot, the top panel shows the evolution of the total luminosity, which is the sum of the luminosity at an optical depth of $\tau = 2/3$ and the power from  positron kinetic energy released by \Co\ decays above the photosphere.  We assume the positrons are able to thermalize and deposit their energy locally while the gamma-rays emitted during the \Ni\ and \Co\ decays escape the wind and the surrounding SN ejecta.  In practice, however, the positron kinetic energy in the optically-thin region is smaller than the luminosity at the photosphere, so this contribution is not significant.  Note that the luminosity from the SN Ia ejecta is not included.    For comparison, the dotted line in the top panels represents the total power from radioactive decays of $0.001 \msol$ of initially pure \Ni\ under ``lab'' conditions.

The effective temperature, derived from the luminosity and radius at the $\tau=2/3$ surface, is shown in the middle panels.  The velocity at the $\tau=2/3$ location is shown as a solid line in the bottom panels, and the mass-averaged velocity in the optically-thin region is shown as a dashed line.

The observed quasi-bolometric late-time light curve of the best observed SN Ia to date, SN 2011fe, is shown as bullets \citep{nuge11,tsve13a,kerz14b,shap16a}.  The  late-time flattening of SN 2011fe and other SNe Ia above simple predictions for \Co\ decay within the SN ejecta has been attributed to light echoes and the decay of long-lived radioactive isotopes such as $^{57}$Co and $^{55}$Fe \citep{seit09b,roep12a,kerz14b,grau16a,shap16a}.    Our models represent an additional contribution to the total luminosity, and thus the observed light curve represents an upper limit to the luminosity for our wind models.

Because the photospheres recede and the models become very hot, peaking at shorter wavelengths than the quasi-bolometric light curves include, it is conceivable that the radiation from the surviving WD companion goes undetected by current observations.  However, the SN ejecta is likely optically thick to these UV photons due to line-blocking even during the nebular phase \citep{li96a}.  Thus, the WD wind's effective temperature while the SN Ia is younger than $\sim 10$ yr is not representative of the expected emission.  Instead, the surviving WD wind's luminosity will be absorbed and re-radiated in the optical by the SN ejecta.   A more quantitative analysis of the expected emission awaits future radiative transfer calculations, but it is likely that the observed late-time light curve of SN 2011fe is indeed an applicable upper limit to the allowed quasi-bolometric luminosity contribution from a surviving WD companion.

It is clear that our models do not fade at the simple \Co\ decay rate at late times.  The delayed \Ni-decays cause the slope of the luminosity evolution to be shallower and, for most models, to flatten to the Eddington-limited value where they remain for years to decades.  This is because the WD envelope regulates itself in such a way that the power released by the  delayed radioactive decays balances gravity at the Eddington limit.  This self-regulation is very similar to the evolution of classical novae, for which the post-nova WD burns hydrogen at the Eddington limit for an extended period of weeks to decades.

Due to this luminosity flattening, it is obvious that the upper limits from \S \ref{sec:Iamotiv} for the captured \Ni\ envelopes on the $0.6$ and $0.9 \msol$ WDs lead to late-time light curves that are too luminous.  Meanwhile, the $0.3 \msol$ WD ejects its much smaller $0.006 \msol$ envelope  more rapidly and continues to fade such that its late-time light curve remains less luminous than the latest observations of SN 2011fe.  However, such a low-mass helium WD cannot be a typical donor for the bulk of SNe Ia due to its long previous main sequence lifetime.

Note that the very sharp drop in luminosity and rise in temperature for the $0.9+0.08 \, M_\odot$ model $\sim 30$ yr after the explosion (green lines in the lower plot of Figure \ref{fig:wdtest}) is somewhat misleading.  The actual evolution of this model as the WD reaches its maximum temperature and then falls onto the WD cooling track occurs at a similar rate as the other models but appears different due to the logarithmic time axis.  This can also be seen in the figures to follow for some of the other models that reach the WD cooling track decades after the explosion.

\begin{figure}
  \centering
  \includegraphics[width=\columnwidth]{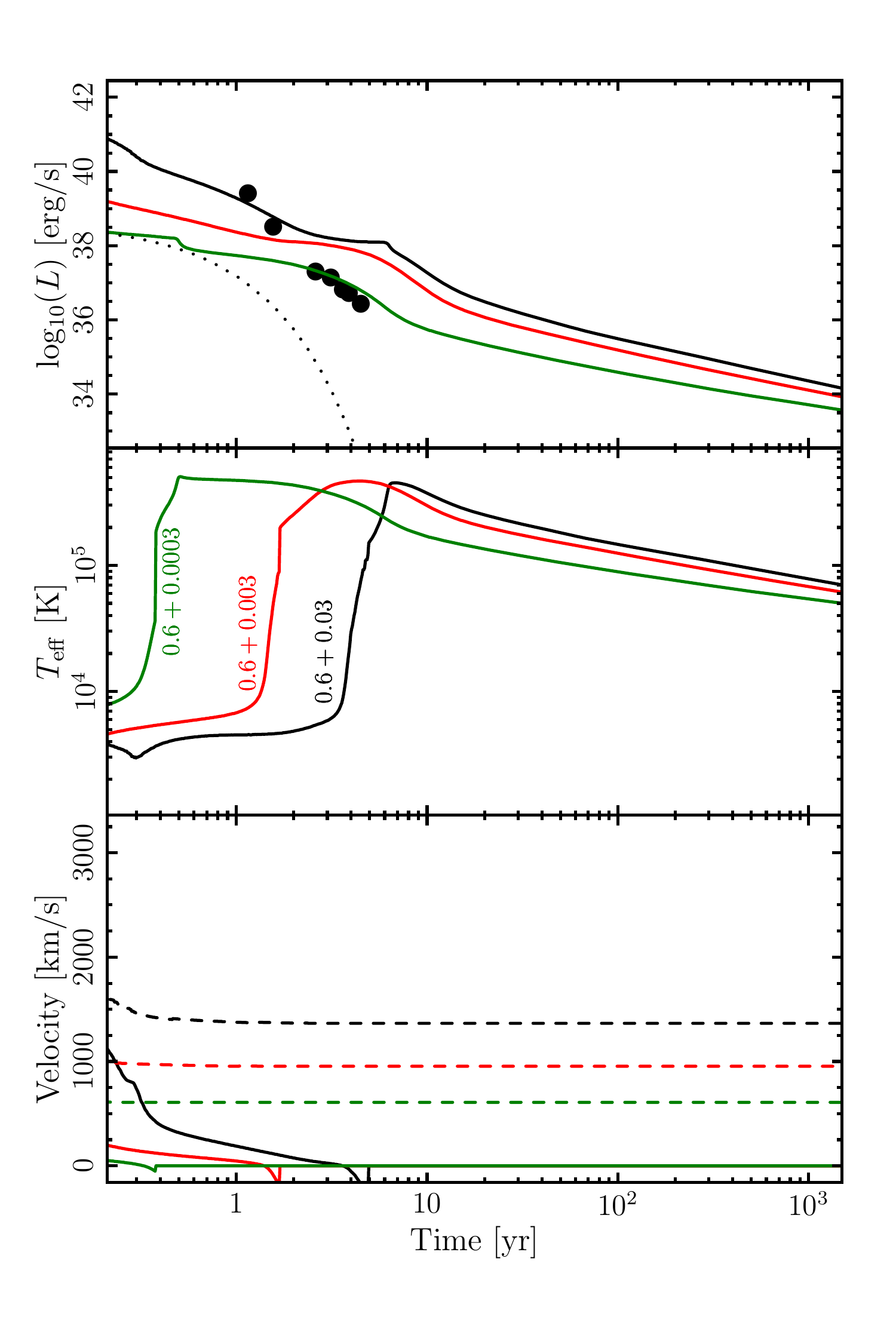}
  \caption{Same as right-hand plot of Figure \ref{fig:wdtest} but for a fixed core mass of $0.6 \msol$ and varying envelope masses of  $0.03$ (black), $0.003$ (red), and $0.0003 \msol$ (green).}
  \label{fig:menvtest}
\end{figure}

Several important assumptions have gone into our modeling of these post-SN Ia winds, and we now demonstrate the effects of relaxing some of these assumptions. In Figure \ref{fig:menvtest}, we show the influence of the choice of the envelope mass.  Black, red, and green lines represent core + envelope masses of $0.6+0.03$, $0.6+0.003$, and $0.6+0.0003 \msol$, respectively.  In order to eject the envelope rapidly enough that the WD can cool and remain less luminous than SN 2011fe at late times, the envelope mass needs to be much smaller than the upper limit of $0.03 \msol$ calculated in \S \ref{sec:Iamotiv}.  This remains a reasonable possibility, because the envelope masses of our fiducial models derived in \S \ref{sec:Iamotiv} are a strict upper limit.  Work done on the ejecta by previously shocked material will reduce the actual mass that remains bound to the surviving WD companion.  Furthermore, although we are assuming the companion WD is not fully disrupted by the onset of the merging process, it will still be significantly expanded when the SN Ia explosion occurs.  Its shallower potential well will also result in a less massive captured envelope.  Future work will use the output of merger simulations to account for these effects and provide more quantitative initial conditions that may yield late-time luminosities within the constraints of SN 2011fe.

\begin{figure}
  \centering
  \includegraphics[width=\columnwidth]{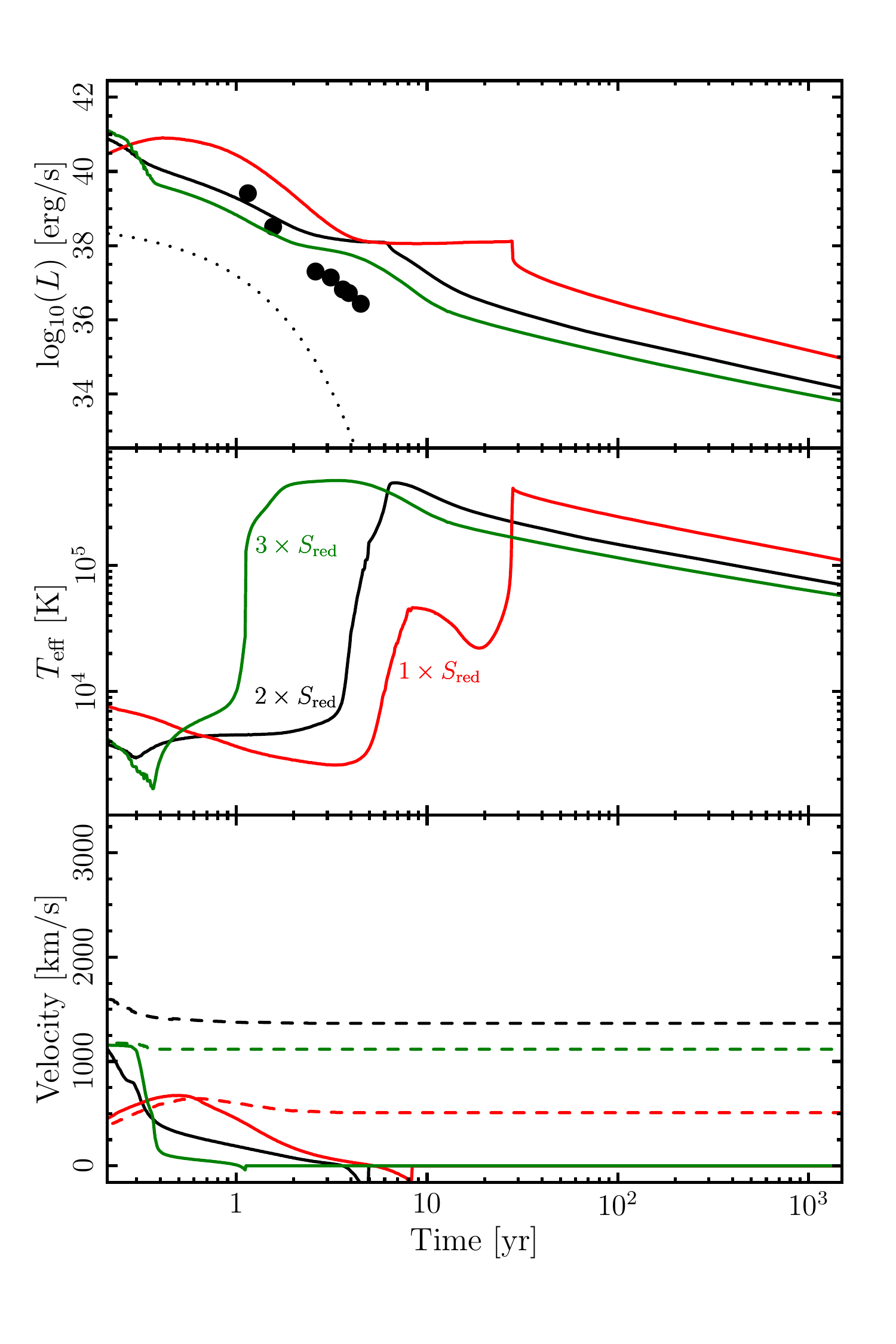}
  \caption{Same as right-hand plot of Figure \ref{fig:wdtest} but for a fixed core + envelope mass of $0.6+0.03 \msol$ and varying initial envelope entropies.  The red model has an initial envelope entropy such that the envelope's integrated internal energy is half its gravitational binding energy.  The black and green models have entropies two times and three times the red model's entropy, respectively.}
  \label{fig:entropytest}
\end{figure}

A related uncertainty is the entropy structure of the captured envelope. It is plausible that our fiducial model is more bound than a realistic simulation would find, and so the fiducial envelope is ejected later than for a realistic initial configuration.  In Figure \ref{fig:entropytest}, we show the effect of changing the entropy of the envelope.  All three models have the same core + envelope mass of $0.6+0.03 \msol$.  The black line, as for all the figures in this subsection showing temporal evolution, is the main fiducial model with an entropy twice that for which the envelope's integrated internal energy is half its integrated gravitational binding energy.  The red and green lines have $0.5$ and $1.5$ times the entropy of the fiducial model, respectively.  As expected, less bound envelopes are ejected sooner, allowing the WD to dim more rapidly.

\begin{figure}
  \centering
  \includegraphics[width=\columnwidth]{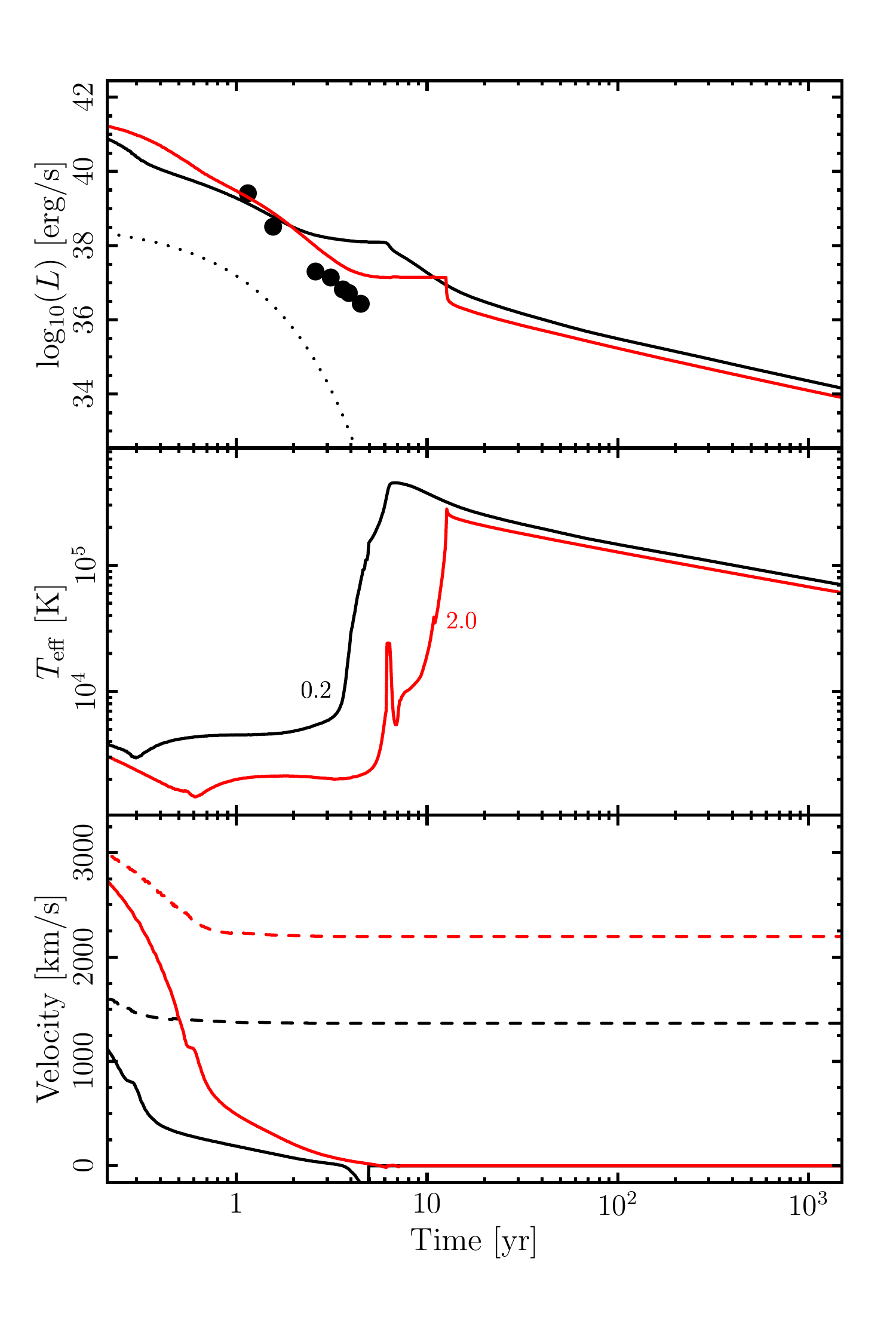}
  \caption{Same as right-hand plot of Figure \ref{fig:wdtest} but for a fixed core + envelope mass of $0.6+0.03 \msol$ and varying opacities.  The fiducial model (black) has a constant opacity of $ \unit[0.2]{cm^2\, g^{-1}}$ in regions less dense than $\unit[1]{g\, cm^{-3}}$, while the red model has a low density opacity of $ \unit[2.0]{cm^2\, g^{-1}}$.}
  \label{fig:kaptest}
\end{figure}

Our final test quantifies the effect of the opacity, as shown in Figure \ref{fig:kaptest}.  The black line is our fiducial model, with a constant opacity $\kappa = \unit[0.2]{cm^2\, g^{-1}}$ in the region with iron group mass fraction $X_{56}>0.1$ and density $\rho < \unit[1]{g\, cm^{-3}}$.  The red line represents a model with a constant opacity $\kappa= \unit[2.0]{cm^2\, g^{-1}}$.  A higher opacity causes the envelope to be ejected somewhat later but also reduces the Eddington limit, so the luminosity flattens at a lower value.  The physical opacity is likely to be higher than our fiducial choice of $\unit[0.2]{cm^2\, g^{-1}}$ because of the large opacity of iron-group elements and effects such as line-broadening in the accelerated wind.  Radiative transfer calculations of post-SN winds will inform a more physical treatment of opacity in future work.

In summary, the main fiducial model shown in Figure \ref{fig:wdtest} is overluminous by several orders of magnitude at the latest observed phases of SN 2011fe, which suggests that if  SNe Ia arise from double WD systems, both WDs are destroyed during the explosion.  However, there are several effects that can reduce the discrepancy between the predicted luminosity from a surviving WD donor's wind and late-time observations.  The largest effect comes from ejecting the \Ni-rich envelope more rapidly, which can in turn be caused by a less massive and less bound envelope.  This is a reasonable expectation since the envelope mass of the fiducial model is an upper limit.  More realistic initial configurations for our wind models from future merger simulations will allow us to use late-time observations to place meaningful constraints on SN Ia progenitor scenarios.


\subsection{Post-SN Iax Winds}

\begin{figure*}
  \centering
  \includegraphics[width=0.45\textwidth]{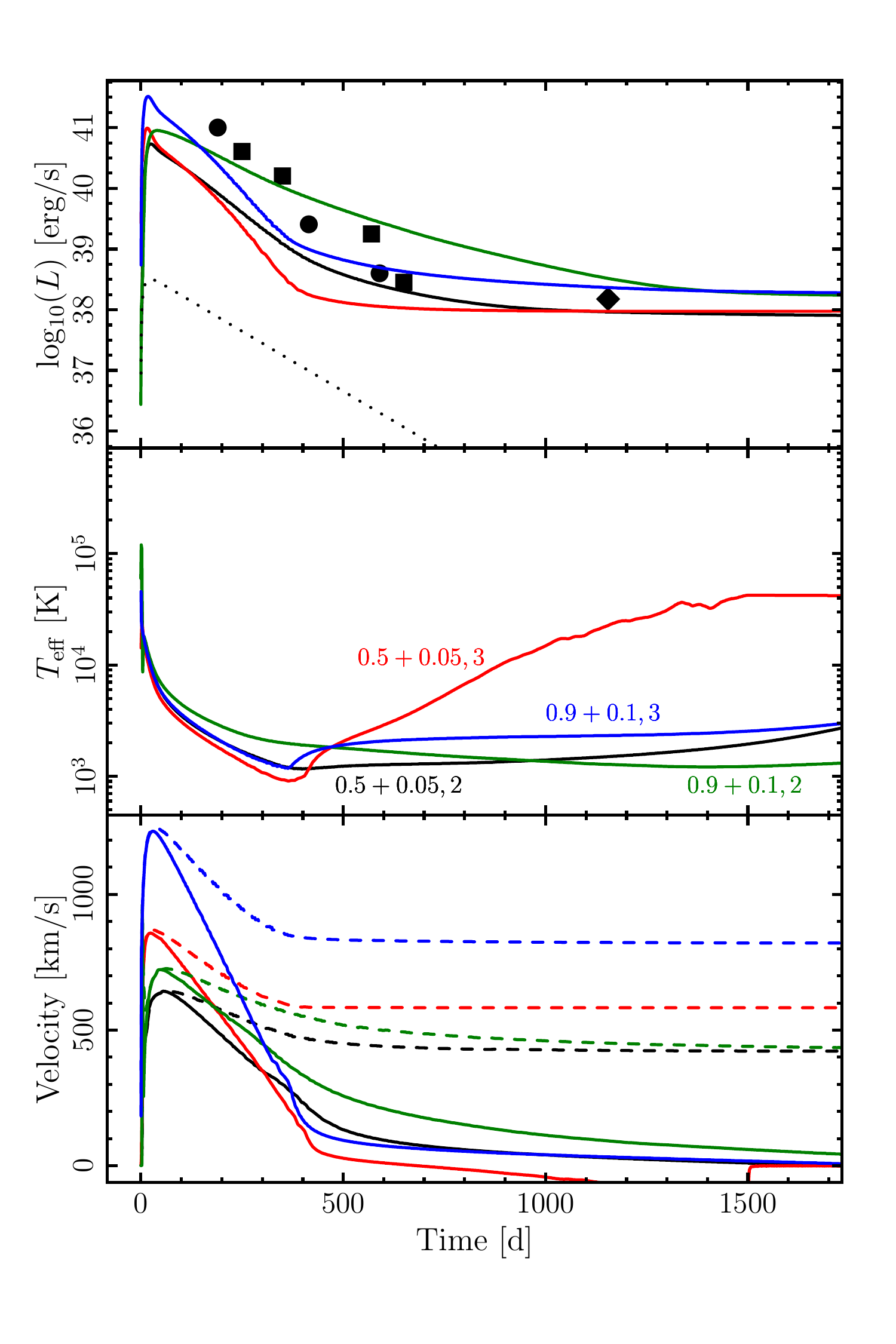}
  \includegraphics[width=0.45\textwidth]{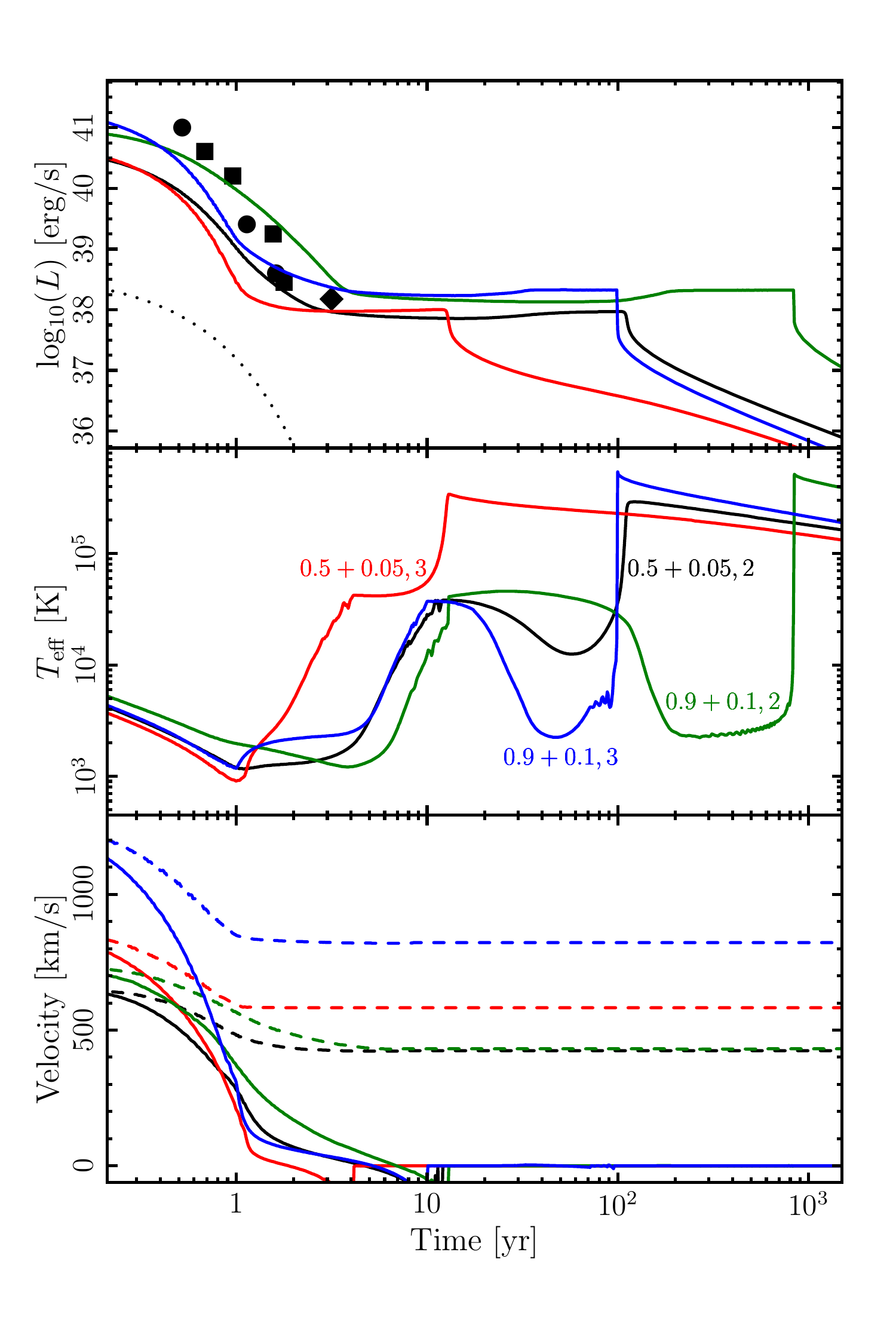}
  \caption{Same as Figure \ref{fig:wdtest} but for our post-SN Iax wind models.  Note that the axis ranges have changed.  Shown are two different mass combinations and two different initial entropies.  Black and red lines represent models with core + envelope masses of $0.5+0.05 \, M_\odot$ and initial \Ni\ mass fractions in the envelope of $0.1$.  Green and blue lines show models with core + envelope masses of $0.9+0.1 \, M_\odot$ and initial \Ni\ mass fractions of $0.2$.  The envelope entropies are two times (black and green) and three times (red and blue) the entropy such that the envelope's internal energy is half its gravitational binding energy.}
  \label{fig:Iax}
\end{figure*}

We now turn to models of winds from bound remnants of SN Iax explosions.  Similar to the post-SN Ia surviving WD donors of the previous subsection, the post-SN Iax surviving WD accretors we model in this subsection will be degenerate cores surrounded by hot \Ni-rich envelopes.  However, as described in \S \ref{sec:Iaxmotiv}, the envelopes will not be pure \Ni, but instead will be a mixture of carbon deflagration ashes with a \Ni\ mass fraction of $0.1-0.2$.

The black and red lines in Figure \ref{fig:Iax} represent core + envelope masses of $0.5+0.05 \msol$ with  initial envelope \Ni\ mass fractions of $0.1$.  The green and blue lines represent masses of $0.9+0.1 \msol$ with initial envelope \Ni\ mass fractions of $0.2$.  The black and green models begin with entropies twice that for which the envelope's integrated internal energy is half its gravitational binding energy; the red and blue models have entropies 50\% higher than the black and green models, respectively.  Models with initial envelope entropies half that of the black and green models were also run.  However, the envelopes were too gravitationally bound for the delayed and diluted  radioactive decays  to launch winds.  Since such a wind is necessary to explain observations of SNe Iax, we do not discuss these models further.

With the exception of the higher mass, lower entropy model (green), our simulations do not significantly exceed the best-observed  late-time SN Iax  bolometric light curves in the literature: SN 2005hk (squares; \citealt{phil07,sahu08a,mccu14b}) 2008A (bullets; \citealt{gane10a,hick12a,mccu14b}), and 2008ha (diamond; \citealt{fole14a}).  The higher mass, higher entropy model (blue) is somewhat marginal, although it is not ruled out given the uncertainties inherent in our simulations.  The photospheric velocities of our models are also in good agreement with the late-time SN Iax velocities reported in the literature \citep{jha06a,mccu14b}.

The large optical depths of the relatively massive SN Iax bound envelopes result in very extended photospheres and low effective temperatures that cool to $ \unit[1000-2000]{K}$ by $\sim \unit[400]{d}$.  The IR detection of SN 2014dt at $ \unit[300-400]{d}$ \citep{fox16a} may be a common feature of SNe Iax and may be due to the low effective temperature of the photosphere, as opposed to, or in addition to, dust formation \citep{fole16a}.

The more massive SN Iax bound envelopes take longer to be ejected as compared to the very low mass envelopes of the SN Ia models favored by the late-time observations of SN 2011fe in the previous subsection.  This is further compounded by the lower initial \Ni\ mass fractions and subsequently lower specific energy release in the SN Iax remnant envelopes.  Thus, it seems to be a robust prediction that the late-time luminosities of these bound remnants will approach the Eddington limit and remain there for years to decades.  In this framework, the $ \unit[1.5 \times 10^{38}]{erg\, s^{-1}}$, $ \unit[2100]{K}$  source detected at the location of SN 2008ha four years after the explosion \citep{fole14a} is consistent with being emission from the bound remnant's wind and not the donor, which should be much hotter if it is a helium star.  We predict a constant bolometric luminosity for SN 2008ha and other old SNe Iax for at least a decade, but the sources will likely become bluer on a shorter timescale of several years.

While the envelope masses in our SN Iax simulations are taken from previous explosion studies and are thus somewhat better motivated than for our SN Ia models, the entropy structures are still a large uncertainty.  More quantitative constraints on SN Iax progenitor models await future studies using hydrodynamic explosion simulations that spatially resolve the bound remnant for more accurate initial conditions.  These future studies should also account for the influence of the surviving donor within the post-SN Iax wind, both as an illuminating source and as a mechanism for shaping and accelerating the wind.


\section{Conclusions}
\label{sec:conc}

In this paper, we have analyzed the radioactive decays of \Ni\ and \Co\ in regimes in which the nuclei may be fully ionized and cannot capture $K$-shell electrons.  We have applied these delayed decays to the evolution of WDs that capture \Ni-rich material following SN explosions.  The WD may be the accreting WD that only partially exploded as a SN Iax, or it may be the former donor to an accreting WD that exploded as a SN Ia.  In both of these situations, a small amount of \Ni-rich material remains bound to the surviving WD at thermodynamic conditions such that no $K$-shell electrons exist for capturing, significantly delaying radioactive decays that power a long-lived wind.  These WD winds will contribute significantly to the observed light curves once the emission from the bulk of the SN ejecta has faded.  We have modeled the appearance of these winds using the stellar evolution code \MESA\ and compared our results to late-time observations of SNe Ia and Iax.

We find that late-time observations of  SN 2011fe strongly constrain the existence of a WD donor that survives a SN Ia.  Our main fiducial model of a $0.6 \msol$ WD that captures a $0.03 \msol$ \Ni-rich envelope is orders of magnitude too bright at the latest observed phases of SN 2011fe.  However, several very important uncertainties remain in our modeling  that allow for the survival of a WD donor.  The most significant uncertainty is the initial conditions we use in our evolutionary calculations.  Beginning with a less massive and less gravitationally bound \Ni-rich envelope results in a shorter elapsed time before the envelope is ejected and the WD dims, allowing it to remain less luminous than SN 2011fe's constraints.  These are reasonable possibilities, and they emphasize the importance of future work that uses the output of double WD merger simulations for more realistic initial conditions.

The uncertainties in our modeling of SN Iax bound remnant winds are somewhat smaller due to relevant previous explosion simulations, although future work will certainly benefit from better initial conditions as well.  Perhaps as a result, most of our fiducial SN Iax wind models do not exceed the late-time luminosities of observed SNe Iax.  In fact, our models match  the latest luminosity, temperature, and velocity measurements of SN 2005hk, SN 2008A, and SN 2008ha reasonably well, giving credence to the near-Chandrasekhar mass, incomplete deflagration model for SNe Iax, with emission initially dominated by the SN ejecta and then by the bound remnant wind at later times.

Even with the large uncertainties in our models, we can make some broad observational predictions that may help to further constrain the progenitor scenarios for SN Ia and SN Iax explosions.  If a WD donor survives a SN Ia and captures a \Ni-rich envelope small enough to remain less luminous than SN 2011fe's latest observations, the WD will fall onto the cooling track within years.  By the age of our nearest Galactic SN Ia remnants ($\ge \unit[400]{yr}$), the WD will be a hot UV source ($T_{\rm eff} \sim \unit[6\E{4}]{K}$) with a luminosity of $\sim L_\odot$, flying away from the center of the explosion at $\unit[1500-2000]{km\, s^{-1}}$.  At a distance of several kpc, this translates to a proper motion of $\sim \unit[0.1]{arcsec\, yr^{-1}}$, observable by the \emph{Hubble Space Telescope} with just a year-long baseline.  We thus highly encourage a UV search for hot, high proper motion stars within Galactic SN Ia remnants.

Our models of SN Iax bound remnant winds strongly suggest that their bolometric luminosities will remain near the Eddington limit for years to decades, with a  somewhat more rapid evolution to higher temperatures.  This provides another motivation for ongoing and future late-time multi-band observations of SNe Iax.  These observations should also extend to the infrared, as the photospheres in some of our models reach temperatures of $ \sim \unit[1000]{K}$ after a year.

This paper represents the first step in a multi-phase modeling effort that will help to constrain thermonuclear SN progenitor scenarios.  Follow-up work will utilize the results of explosion simulations for better initial conditions, which will be particularly crucial for narrowing down our predictions of post-SN Ia surviving WD donors.  Another important avenue of research will be radiative transfer calculations through the post-SN winds.  These will allow comparisons to the peculiar late-time photospheric spectra of SNe Iax.  Furthermore, our newly-proposed long-lived radiation source at the center of SN Ia ejecta may have some connection to some of the persistent puzzles of late-time SN Ia spectra, including the existence of bimodal line profiles \citep{dong15a}.  This previously unconsidered source of radiation will also affect the ionization balance within the SN Ia ejecta, which may alter inferred mass estimates of stable $^{54}$Fe and $^{58}$Ni \citep{mazz15a}.  Additionally, the possible presence of recently launched \Ni\ in the wind that has not yet decayed may explain the detection of Ni in nebular spectra, which has previously been interpreted as evidence for the production of stable $^{58}$Ni in SNe Ia \citep{mazz15a,frie16a}.


\acknowledgments

We thank Lars Bildsten, Ryan Foley, Claes Fransson, Dan Kasen, Wolfgang Kerzendorf, Alison Miller, R\"{u}diger Pakmor, Eliot Quataert, and Stuart Sim for helpful
discussions, and Lin-Manuel Miranda and Leslie Odom, Jr., for inspiring the title.  KJS is supported by NASA through the Astrophysics Theory
Program (NNX15AB16G) and by the Research Corporation for Science
Advancement through a ``Scialog: Time Domain Astrophysics'' award.
JS is supported by NASA through Hubble Fellowship
grant HST-HF2-51382.001-A awarded by the Space Telescope Science
Institute, which is operated by the Association of Universities for
Research in Astronomy, Inc., for NASA, under contract NAS5-26555.
This research used the Savio computational cluster resource provided
by the Berkeley Research Computing program at the University of
California, Berkeley (supported by the UC Berkeley Chancellor, Vice
Chancellor of Research, and Office of the CIO).
  
\software{MESA}



\appendix

We use the stellar evolution code \MESA\ to evolve our post-SN wind models.  We construct initial models with degenerate cores and hot \Ni-rich envelopes as outlined in \S \ref{sec:sim}.  We then begin the hydrodynamic evolution of the model with opacity and energy routines as described in \S \ref{sec:sim} and the following inlist:

\

\texttt{\&star\_job}

\

\texttt{change\_net = .true.}

\texttt{new\_net\_name = `basic\_plus\_2856.net' }

\texttt{auto\_extend\_net = .false.}

\texttt{change\_v\_flag = .true.}

\texttt{new\_v\_flag = .true.}

\texttt{relax\_tau\_factor = .true.}

\texttt{relax\_to\_this\_tau\_factor = 1d-4}

\

\texttt{\&controls}

\

\texttt{min\_timestep\_limit = 1d-99}

\texttt{logQ\_min\_limit = -99}

\texttt{logQ\_limit = 99}

\texttt{use\_Type2\_opacities = .true.}

\texttt{Zbase = 0.0133d0}

\texttt{use\_other\_kap = .true.}

\texttt{use\_other\_energy = .true.}

\texttt{max\_allowed\_nz = 99999999}

\texttt{min\_dq = 1d-14}

\texttt{max\_surface\_cell\_dq = 1d99}

\texttt{mesh\_min\_dlnR = 1d-7}

\texttt{merge\_if\_dlnR\_too\_small = .true.}

\texttt{mesh\_adjust\_get\_T\_from\_E = .false.}

\texttt{MLT\_option = `none'}

\texttt{use\_ODE\_var\_eqn\_pairing = .true.}

\texttt{use\_dvdt\_form\_of\_momentum\_eqn = .true.}

\texttt{use\_dPrad\_dm\_form\_of\_T\_gradient\_eqn = .true.}

\texttt{use\_momentum\_outer\_BC = .true.}

\texttt{use\_compression\_outer\_BC = .true.}

\texttt{use\_zero\_dLdm\_outer\_BC = .true.}

\texttt{use\_artificial\_viscosity = .true.}

\

Once a wind has been established or three simulation days have passed, convection with velocities limited to the sound speed is enabled and two outer boundary conditions are changed to make the wind more stable:

\

\texttt{\&controls}

\

\texttt{MLT\_option = `Cox'}

\texttt{min\_T\_for\_acceleration\_limited\_conv\_velocity = 0}

\texttt{mlt\_accel\_g\_theta = 1}

\texttt{max\_conv\_vel\_div\_csound = 1}

\texttt{use\_zero\_Pgas\_outer\_BC = .true.}

\texttt{use\_zero\_dLdm\_outer\_BC = .false.}

\

Finally, when the WD ceases launching the wind and material begins to fall back onto its surface, the unbound material is rapidly removed using flags that force the material to be removed from zones near the surface.

\

\texttt{\&controls}

\

\texttt{mass\_change = -1d2}

\texttt{min\_q\_for\_k\_below\_const\_q = 0.999999999999999}

\texttt{min\_q\_for\_k\_const\_mass = 0.999999999999999}



\end{document}